\documentclass[10pt,letterpaper]{article}

\usepackage{cogsci}
\usepackage{pslatex}
\usepackage{amsmath,dsfont,amsthm,amssymb}
\usepackage{tikz}
\usepackage{tkz-graph}
\usepackage{apacite}
\usepackage[colorinlistoftodos]{todonotes}
\usepackage[skip=4pt]{caption}
\usepackage{bbm}

\title{A Formal Approach to Modeling the Cost of Cognitive Control}
 
\author{
Kayhan {\"O}zcimder $^{1,2,*}$, Biswadip Dey$^{2,*}$, Sebastian Musslick$^{1,*}$, \\ 
Giovanni Petri$^3$, Nesreen K. Ahmed$^4$, Theodore L. Willke$^4$, Jonathan D. Cohen$^1$ \\
  $^1$Princeton Neuroscience Institute, Princeton University, Princeton, NJ 08540, USA. \\
  $^2$Department of Mechanical and Aerospace Engineering, Princeton University, Princeton, NJ 08544, USA. \\
  $^3$ ISI Foundation, Via Alassio 11/c, 10126 Torino, Italy.\\
  $^4$Intel Labs, Santa Clara, CA 95054, USA.\\
  $^*$Equal Contribution, Corresponding Author: {\it ozcimder@princeton.edu}
}

\DeclareMathOperator{\erf}{erf}
\begin{document}

\maketitle

\begin{abstract}
This paper introduces a formal method to model the level of demand on control when executing cognitive processes. The cost of cognitive control is parsed into an \emph{intensity cost} which encapsulates how much additional input information is required so as to get the specified response, and an \emph{interaction cost} which encapsulates the level of interference between individual processes in a network. We develop a formal relationship between the probability of successful execution of desired processes and the control signals (additive control biases). This relationship is also used to specify optimal control policies to achieve a desired probability of activation for processes. We observe that there are boundary cases when finding such control policies which leads us to introduce the interaction cost. We show that the interaction cost is influenced by the relative strengths of individual processes, as well as the directionality of the underlying competition between processes.
\end{abstract}

%
%
%%%%%%%%%%%%%%%%%%%%%%%%%%%%%%%%%%%%%%%%%%%%%%%%%%%%%%%%%%%%%%%%%%%%%%%%%%%%%%%%%%%%%%%%%%%%%%%%%%%%%%%%
\section{Introduction}
\label{sec:Introduction}
%%%%%%%%%%%%%%%%%%%%%%%%%%%%%%%%%%%%%%%%%%%%%%%%%%%%%%%%%%%%%%%%%%%%%%%%%%%%%%%%%%%%%%%%%%%%%%%%%%%%%%%%
%%%%%%%%%%%%%%%%%%%%%%%%%%%%%%%%%%%%%%%%%%%%%%%%%%%%%%%%%%%%%%%%%%%%%%%%%%%%%%%%%%%%%%%%%%%%%%%%%%%%%%%%
%
A long standing focus in cognitive research has been towards understanding the ability to execute tasks/processes\footnote{A task/process/input-output mapping is defined as a unique mapping from all possible vectors in the input subspace to corresponding vectors in the output subspace, that is independent of the mappings for all other combinations of input and output components in the network.} that demand \emph{cognitive control}. In this context, cognitive control is defined as the set of mechanisms required to pursue a goal, especially when distraction or strong competing responses (interferences) must be overcome \cite{posnerr,shiffrin1977controlled,cohen1990control}. Earlier work \cite{posnerr, shiffrin1977controlled, cohen1990control, COGS:COGS12126} has argued that the processes demanding control can be distinguished from \emph{automatic processes} in terms of the strength of the associations in the pathways underlying processing: automatic processes are characterized by pathways with associations strong enough to resist interference from competing processes, whereas controlled processes are weaker, and therefore rely on input from control mechanisms to support their execution against interference.

Another longstanding observation is that the allocation of cognitive control is costly (often discussed in terms of ``mental effort" \cite{posnerr, Botvinick_Braver_2015, Shenav_PrePrint_ARP}). This cost has been interpreted in physical terms (such as metabolic demands \cite{muraven1998self}) or in terms of an opportunity cost reflecting the allocation of a limited resource \cite{kurzban2013opportunity}. Elsewhere \cite{Feng_et_al_2014, CogSci_2016}, we have proposed that limitations in the capacity for control-dependent processing reflect the \emph{purpose} of control —to diminish interference —rather than any intrinsic limitation in the mechanism responsible for control. This view suggests that the architecture of the processing system as a whole constrains the opportunities for control-dependent processing, resulting in opportunity costs associated with allocating control to any particular task(s). 

Here, we build on a closely related proposal, by \citeA{koechlin2007information}, to define the cost of control in terms of internal representational requirements to insure that a given stimulus (or a set of stimuli) produces the desired response (or a set of responses), given the intrinsic architecture of the system. Their work focused on a single task. Here, we extend this to consider an arbitrary number of tasks and thus accommodate their possibility for, and costs of, multitasking (i.e.parallel processing of task pathways). To do so, we follow the framework proposed by \citeA{shenhav2013expected} that distinguishes two components of control signals: \emph{intensity} and \emph{identity}. Specifically, \citeA{shenhav2013expected} defined the intensity of a control signal as the strength of the signal needed to insure performance of a particular task, and the identity as which control signal should be selected to achieve a desired objective given environmental conditions. Here we build on that distinction to define two corresponding components of the cognitive control costs - a cost associated with intensity, and a cost associated with interaction. Furthermore, we define the interaction cost to capture the level of interference between the processes in a network.

In this paper, we begin by introducing formal constructs for intensity and interaction costs by using the graph theoretic representation of a neural network and terms/notions adopted from probability theory. We describe an intensity cost that represents the control signals (as biases infused into a neural network), above and beyond the specified strength of the signal (stimulus) itself. This is achieved by developing a formal relationship between the probability of successful execution of desired processes and the control signals. 
In turn, this defines an optimization problem, which can then be solved to find optimal control signals that achieve a specified activation for desired processes. However, we observe that there are boundary cases in which this optimization problem can not be solved. These cases reflect situations in which the simultaneous execution of the processes is not feasible due to interference. Hence, interaction cost analysis motivates an additional investigation towards finding a proper metric that continuously measures the level of interference between processes. To achieve this we introduce the definition of interaction cost associated with process mappings in a network configuration. Specifically, it measures the level of interference introduced by competing processes that interfere with the tasks of interest. In their study, \citeA{koechlin2007information} have already used information theoretic terms to measure cognitive control. However, in order to apply these metrics to neural networks while considering parallelism, we augment some of these measures, and through simulations we will demonstrate how interaction cost can be used to predict interference in neural network architectures. Finally, we will discuss general research directions revealed by the analysis presented here.
%
%
%

%
%
%%%%%%%%%%%%%%%%%%%%%%%%%%%%%%%%%%%%%%%%%%%%%%%%%%%%%%%%%%%%%%%%%%%%%%%%%%%%%%%%%%%%%%%%%%%%%%%%%%%%%%%%
\section{Intensity: the cost of control}
%\textcolor{pink}{What about changing this to: "Intensity: the cost of control"}
%%%%%%%%%%%%%%%%%%%%%%%%%%%%%%%%%%%%%%%%%%%%%%%%%%%%%%%%%%%%%%%%%%%%%%%%%%%%%%%%%%%%%%%%%%%%%%%%%%%%%%%%
%%%%%%%%%%%%%%%%%%%%%%%%%%%%%%%%%%%%%%%%%%%%%%%%%%%%%%%%%%%%%%%%%%%%%%%%%%%%%%%%%%%%%%%%%%%%%%%%%%%%%%%%
%
Cognitive control is defined as the underlying mechanism that biases the processing of a task in order to maximize the reward \cite{Botvinick_etal_2001, Botvinick_etal_2004, Bogacz_OptDecMak, Botvinick2007}. Here, we adapt the notion of intensity cost from \citeA{shenhav2013expected} as a function of the amount of control bias that cognitive control applies to the system. However, \citeA{shenhav2013expected} described this function in qualitative rather than quantitative terms. In this work, we provide an explicit characterization of the cost of cognitive effort in terms of a set of physically meaningful parameters, which allow the manipulation of the response of a cognitive architecture.
\begin{figure}[tb]
\begin{center}
  \includegraphics[width=0.35\textwidth]{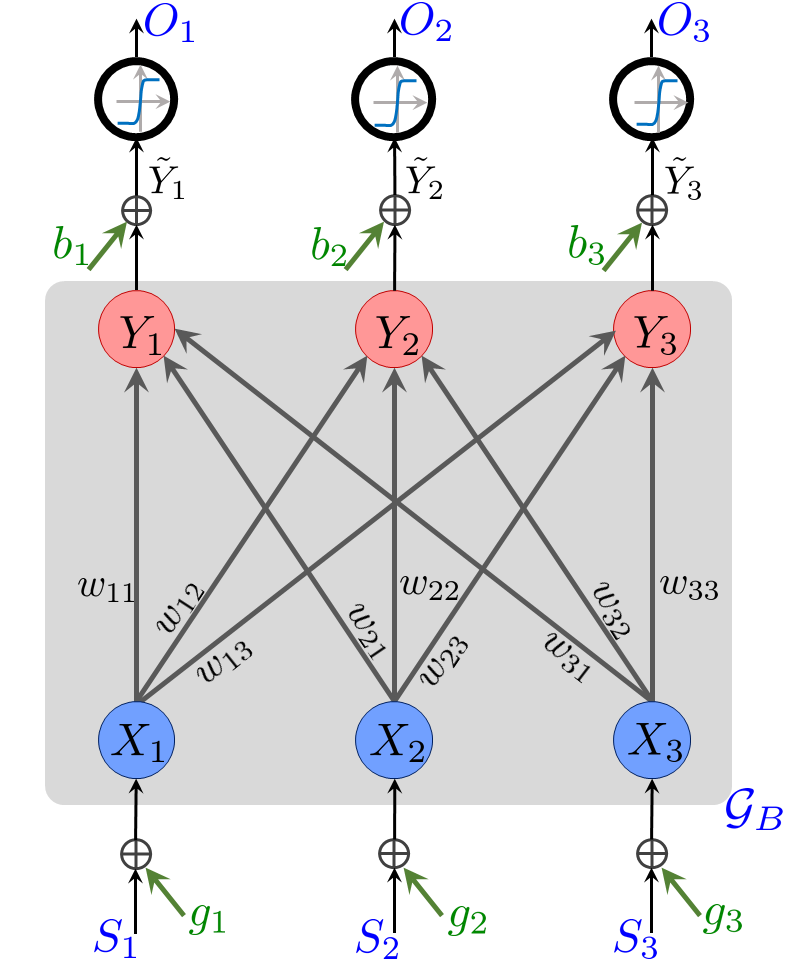}
  \caption{\small{Illustration of a single-layered, feed-forward network with 3-input and 3-output layer components, wherein the individual features are scalar in nature.}}
  \label{SLFF_Net_3}
\end{center}
\vspace{-2.0em}
\end{figure}

Following earlier works \cite{Feng_et_al_2014, CogSci_2016}, we consider a single-layered, feed-forward network with $N$ input and $M$ output layer components to formalize the notion of intensity cost in a cognitive control context (Fig.~\ref{SLFF_Net_3} shows a simple example of such a network). In this framework, each component represents an input/stimulus or output/response dimension (vector subspace), and the connection from an input to an output component constitutes the processing pathway for a given task. This allows us to define an abstraction of the network as a directed bipartite graph $\mathcal{G}_B = (\mathcal{V},\mathcal{E})$, wherein the set of vertices $\mathcal{V}$ can be partitioned into two disjoint sets $\mathcal{V}_{in}$ and $\mathcal{V}_{out}$, representing the input and output layer components respectively. Moreover, a directed edge $(i,j) \in \mathcal{E} \subseteq \mathcal{V}_{in} \times \mathcal{V}_{out}$ represents a connection from the vertex $i$ in the input layer to vertex $j$ in the output layer (i.e., a task). In this setting, we represent the processing pathway by introducing a weight matrix $W$ with elements $w_{ij}$. As we will see later, this abstraction plays an important role in formalizing the interaction cost of cognitive control.

In this setting, we assume that control signals bias the processing of a stimulus towards a specified response at two different levels, i.e. $g_i$ and $b_j$, which we refer to as pre-interaction and post-interaction control biases, respectively. This complies with early computational models of cognitive control in which control signals act as an increase in gain of non-linear processing units \cite{cohen1990control,Botvinick_etal_2001} and allows us to treat such control biases as key contributing factors towards the intensity cost for cognitive control. It is worth noting here that for simplicity the only sources of nonlinearity in this setting are the logistic\footnote{Although we are restricting ourselves to logistic functions with unit steepness, in a more general setting one can use the steepness as another design parameter.} activation functions which act upon the linearized output vector $\tilde{\mathbf{y}} = [\tilde{Y}_1, \tilde{Y}_2, \ldots, \tilde{Y}_M]$. Without loss of generality, in what follows we consider the individual features to be scalar, and carry out a formal investigation on how these control biases $\mathbf{g} = [g_1,\ldots,g_N]$ and $\mathbf{b} = [b_1,\ldots,b_M]$ influence the response from this cognitive architecture. In our formulation, the corresponding magnitude, i.e. $\|\mathbf{g}\|^2+\|\mathbf{b}\|^2$, can be treated as a measure of control intensity applied to the system, and therefore the cost for cognitive control.

We begin our analysis by assuming the vector of features $\mathbf{s} = [S_1,S_2,\ldots,S_N]$ to be an $N$-dimensional multivariate Gaussian random variable with mean $\mu^S$ and covariance $\Sigma^S$. (The assumption of Gaussianity is motivated by the technical tractability). With this assumption, the vector $[X_1, X_2, \ldots, X_N]$ becomes an $N$-dimensional multivariate Gaussian random variable with a shifted mean and same covariance. Furthermore, as all the transformations (before the nonlinear logistic activation function) are linear in nature, $[Y_1,Y_2,\ldots,Y_N]$ also remains a multivariate Gaussian random variable whose mean and covariance are given by $\mu^Y = W(\mu^S+\mathbf{g})$ and $\Sigma^Y =  W \Sigma^S W^T$, respectively. Similarly, the vector of linearized outputs $[\tilde{Y}_1, \tilde{Y}_2, \ldots, \tilde{Y}_N]$ is also a multivariate Gaussian, with a shifted mean and the same covariance.

Hence, each individual linearized output $\tilde{Y}_i$ is itself a Gaussian random variable with 
\begin{displaymath}
\begin{aligned}
\textrm{mean:} && \mu^{\tilde{Y}}_i &= \sum\limits_{j=1}^N w_{ji}(\mu_j + g_j) + b_i
\\
\textrm{and variance:} && \sigma^{\tilde{Y}}_{ii} &= \sum\limits_{j=1}^N \sum\limits_{k=1}^N w_{ji}w_{ki}\sigma_{kj},
\end{aligned}
\end{displaymath}
where $\mu_j$ is the mean of stimulus $S_k$ and $\sigma_{kj}$ is the covariance between stimuli $S_k$ and $S_j$. As a consequence, the corresponding output (response) will have a logit-normal distribution, and this leads us to our key result in this section.

As outlined by \citeA{shenhav2013expected}, the response $O_i$ should overcome a specified threshold in order to execute the corresponding process (task). Then, by letting $\alpha_i \in (0,1)$ represent this activation threshold associated with output $O_i$, the corresponding probability of task execution (probability of the output $O_i$ surpassing the threshold $\alpha_i$) is expressed as
\begin{displaymath}
P[O_i \geq \alpha_i] = \frac{1}{2} - \frac{1}{2}\underbrace{\erf\left(\frac{\log\left(\frac{\alpha_i}{1-\alpha_i}\right) - b_i - \sum\limits_{j=1}^N w_{ji}(\mu_j + g_j)}{\sqrt{2\sum\limits_{j=1}^N \sum\limits_{k=1}^N w_{ji}w_{ki}\sigma_{kj}}}\right)}_{f(\alpha_i, b_i, w, \mu, \mathbf{g}, \Sigma^S)}.
\end{displaymath}
Here we have exploited the monotonicity of the logistic function to compute its inverse. Then the result follows from the cumulative distribution function of $\tilde{Y}_i$.

We characterized the activation probability of a given network in terms of the pre-interaction and post-interaction control biases. This is crucial because it provides new directions to incorporate the cost of control into the design of a cognitive network architecture. For example, the problem of allocating a limited amount of cognitive control into different components of the network to maximize the associated probability of activation can be formulated as an optimization problem in which the goal becomes minimizing $f(\alpha_i, b_i, w, \mu, \mathbf{g}, \Sigma^S)$ over $\mathbf{g}$ and $\mathbf{b}$ subject to the constraints $\sum\limits_{i=1}^N g^2_i \leq C_g$ and $\sum\limits_{i=1}^M b^2_i \leq C_b$, where $C_g$ and $C_b$ define the maximum amount of control that can be applied. Alternatively, in this setting, we can also approach the problem of minimizing the cost of control, while still maintaining a desired value of probability of activation.

One can consider the joint distribution of the processes of interest to incorporate the effects of interaction between tasks. To be consistent with \cite{Feng_et_al_2014, CogSci_2016}, it is reasonable to begin with a focus narrowed to the situation where the choice of interaction weights and the prior distribution of the stimuli render the interactions undesirable. Then the effect of multitasking can measured by introducing a suitable distance metric (for example, the Kullback-Leibler divergence \cite{ortega2013thermodynamics}) between the joint distributions of relevant processes and the product of corresponding marginals, and one can attempt to minimize this distance at the expense of a limited amount of cognitive control. However, as one might expect, this optimization problem can have an empty solution set under certain values of the interaction weights and activation thresholds, meaning that certain network configurations strictly prohibit successful multitasking performance \cite{CogSci_2016}. Before approaching this computation in detail, it would be beneficial to investigate how the interaction structure influences the solution space, and that leads us to our next section wherein we introduce the notion of interaction cost.
%
%
%

%
%
%%%%%%%%%%%%%%%%%%%%%%%%%%%%%%%%%%%%%%%%%%%%%%%%%%%%%%%%%%%%%%%%%%%%%%%%%%%%%%%%%%%%%%%%%%%%%%%%%%%%%%%%
\section{Interaction: the cost of mapping}
\begin{figure}[t]
\begin{center}
  \includegraphics[width=0.45\textwidth]{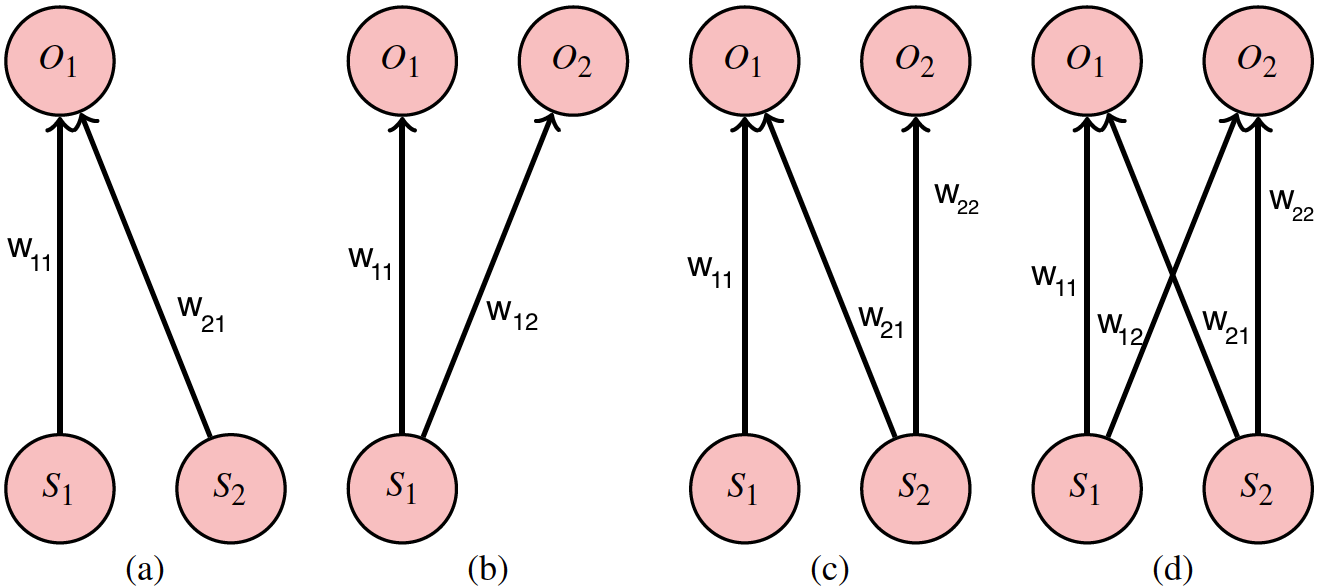}
  \caption{\small{The illustration of convergent, divergent, asymmetric, and symmetric interference.}}
  \label{indirect}
\end{center}
\vspace{-2.0em}
\end{figure}

%\textcolor{pink}{what about? ``Interaction: the cost of mapping''}
%%%%%%%%%%%%%%%%%%%%%%%%%%%%%%%%%%%%%%%%%%%%%%%%%%%%%%%%%%%%%%%%%%%%%%%%%%%%%%%%%%%%%%%%%%%%%%%%%%%%%%%%
%%%%%%%%%%%%%%%%%%%%%%%%%%%%%%%%%%%%%%%%%%%%%%%%%%%%%%%%%%%%%%%%%%%%%%%%%%%%%%%%%%%%%%%%%%%%%%%%%%%%%%%%
%
In this section, we will introduce a detailed formalism of the interaction cost associated with process mappings in a network configuration to accommodate the possibility for multitasking. In our earlier work \cite{CogSci_2016}, we have formalized three distinct types of interference (Fig.~\ref{indirect}).

\emph{Convergent interference} (Fig.~\ref{indirect}a) occurs when two inputs/stimuli (e.g. $S_1$ and $S_2$) compete to determine a common output (e.g. $O_1$). We also consider \emph{divergent interference} in our analysis (Fig.~\ref{indirect}b). Although this does not pose an impediment to performance, i.e. it is possible to generate two distinct outputs (e.g. $O_1$ and $O_2$) to the same input (e.g. $S_1$), it represents a restriction on the number of independent stimuli (and therefore the number of tasks) that the system can process at once, and thus was treated formally as a type of interference due to this dependency in our analysis of parallel processing capability. Finally, we consider a third, \emph{indirect interference} that supervenes on the first two (shown in Fig.~\ref{indirect}c and Fig.~\ref{indirect}d). In this case, the two tasks with strengths $w_{11}$ and $w_{22}$ in question do not directly interfere with one another. However, their simultaneous execution would necessarily engage a third task with strength $w_{21}$ (also possibly a fourth task with strength $w_{12}$) that would produce interference in output $O_1$ (and $O_2$). While \citeA{CogSci_2016} treated these three types of interference identically in terms of their effect on the overall parallel processing capability of a network, the proposed interaction cost will also distinguish between these three types of interference.

In interaction cost analysis, we will assume that a stimulus is of value $1$ when it is active, and $0$ otherwise. Moreover, to increase tractability, we will consider linear activation at the output level, which also implies that without loss of generality the pre- and post-interaction biases can be assumed to be zero. A more detailed version of the interaction cost analysis, involving the strength of stimuli, as well as the nonlinear activation function, will be discussed in subsequent publications.

To introduce the interaction cost, we take an approach similar to the one adopted by \citeA{koechlin2007information}. In their work, \citeA{koechlin2007information} proposed a metric for selecting a single action among multiple alternatives. Here, we will refine this metric to introduce the interaction cost for neural network architectures. Towards this objective, we first leverage the assumptions discussed earlier in the section, and abstract out the network configurations presented in Fig.~\ref{indirect} from the network shown in Fig.~\ref{SLFF_Net_3}. We also assume that the strength of a task $ij$ from stimulus $S_i$ to output $O_j$ is represented by its non-negative weight $w_{ij} \geq 0$. 

Let us first consider the case shown in Fig.~\ref{indirect}a. It is obvious that the response in the output component $O_j$ is completely determined by the stimulus if either $S_1$ or $S_2$ is activated in the network (executing a single process). However, activating both stimuli $S_1$ and $S_2$ simultaneously creates a conflict, since the output can not respond to two distinct stimuli simultaneously (as the activations are linear, the network will always have a response in the output level). In order to measure the level of this competition between stimuli, we define a random variable $a_1$ associated with the output $O_1$ such that $a_1 \in \{1,2\}$ (Fig.~\ref{indirect}a). This implies that $a_1=1$ or $a_1=2$ when the output $O_1$ is driven completely by $S_1$ or $S_2$, respectively. Since a stronger task will have a higher probability of being selected to generate the response, we consider the relative strengths of the task pathways (with associated strengths $w_{11}$ and $w_{21}$) in order to define the probability of the possible outcomes, i.e. the probability of $a_1=1$ and $a_1=2$ when both $S_1$ and $S_2$ activated. Hence, we compute the probability as
\begin{displaymath}
P[a_1=1] = \frac{w_{11}}{w_{11}+w_{21}},
\quad \textrm{and,} \quad
P[a_1=2] = \frac{w_{21}}{w_{11}+w_{21}}.
\end{displaymath} 

Next, we extend this framework to consider networks with $N$ stimuli and $M$ outputs, wherein an output $O_j$, $j \in \{1, \ldots, M\}$ responds to a set of stimuli. Let us assume that there are $n \leq N$ incoming edges to a particular output $O_j$, and each edge is originated from a distinct stimulus $S_i$, $i=i_1, \ldots, i_n$, where $i_k \in \{1, \ldots, N\}$. Then the probability of the event that output $O_j$ is responding to stimulus $S_i$ is given by,
\begin{equation}
\label{proba}
P[a_j=i] = \frac{w_{ij} \mathbbm{1}(S_i)}{\sum\limits_{k = 1}^n w_{i_kj}\mathbbm{1}(S_{i_k})},
\end{equation}
where $\mathbbm{1}(S_i)$ is the indicator function that represents the activation of stimulus $S_i$ such that $\mathbbm{1}(S_i)=1$ if stimulus $S_i$ is active and $\mathbbm{1}(S_i)=0$, otherwise. Then, by building upon the ideas proposed by \cite{koechlin2007information}, we define the interaction cost as
\begin{equation}
\Psi(a_j=i) = -\log(P[a_j=i]),
\end{equation}
where the logarithm is with respect to base 2.

Equation~\ref{proba} implies the $P[a_j=i] = 1$ when only the relevant stimulus $S_i$ associated with task $ij$ is activated in the network. Hence the interaction cost is computed as $\Psi(a_j=i) = 0$ which implies that there is no interaction cost. Moreover, when multiple processes are competing due to the activation of multiple stimuli, $P[a_j=i] \rightarrow 0$ as the competition increases, and as a consequence the interaction cost $\Psi(a_j=i) \rightarrow \infty$.

We further extend equation \ref{proba}, to encapsulate the probability associated with parallel processing of task pathways in the network. Thus, we introduce the joint probability of distinct output components responding to a set of stimuli. For instance, let us consider the parallel processing of tasks with strength $w_{11}$ and $w_{21}$ in Fig.~\ref{indirect}a, and calculate $P[a_1=1, a_1=2]$. This is the probability of output component $O_1$ responding to both $S_1$ and $S_2$, and by definition we know that this probability is zero. For the case illustrated in Fig.~\ref{indirect}b, the joint probability $P[a_1=1, a_2=1] = 1$ since activation of $S_1$ will activate both processes with strengths $w_{11}$ and $w_{12}$, and there is no competition in outputs $O_1$ and $O_2$. This result is parallel to the observation made by \cite{CogSci_2016}, who stated that divergent interference is not actually an interference but a dependency on the stimuli.

Now let us consider the case introduced in Fig.~\ref{indirect}c which can be thought of as the composition of the two cases presented in Fig.~\ref{indirect}a-b. We compute the interaction cost of parallel processing the tasks with strengths $w_{11}$ and $w_{22}$. This requires simultaneous activation of $S_1$ and $S_2$, which indirectly engages the task with strength $w_{21}$, and initiates a competition in the output $O_1$. Thus, the interaction cost of parallelism between tasks represented by $w_{11}$ and $w_{22}$ is given by
\begin{align*}
\Psi_1(a_1=1,a_2=2)
&= -\log(P[a_2=2] \cdot P[a_1=1|a_2=2])
\\
&= -\log\left(1 \cdot \frac{w_{11}}{w_{11}+w_{21}}\right).
\end{align*}
Here $P[a_2=2]=1$ since task with weight $w_{22}$ is not competing with any other process in the output $O_2$. The competition, however, takes place in $O_1$, and the interaction cost associated with $w_{11}$ for this case has already been computed when we discussed the case in Fig.~\ref{indirect}a. 

In a similar way, we can compute the interaction cost of parallelism between tasks represented by $w_{11}$ and $w_{22}$ in Fig.~\ref{indirect}d, and we have
\begin{align*}
\Psi_2(a_1=1,a_2=2)
&= -\log(P[a_2=2] \cdot P[a_1=1|a_2=2])
\\
&= -\log\left(\frac{w_{11}}{w_{11}+w_{21}} \cdot \frac{w_{22}}{w_{22}+w_{12}}\right).
\end{align*}
In this case, simultaneous activation of $S_1$ and $S_2$ causes competition in both outputs $O_1$ and $O_2$. Thus, by revealing further insight about the strength and directionality of interference, the interaction cost serves as an extension of the interference definition presented by \cite{CogSci_2016}. For instance, for the same values of $w_{11}, w_{21}, w_{22} \geq 0$ in both configurations in Fig.~\ref{indirect}c-d, and given $w_{12} \geq 0$ for the configuration in Fig.~\ref{indirect}d, it is obvious that 
\begin{displaymath}
\Psi_1(a_1=1,a_2=2) \leq \Psi_2(a_1=1,a_2=2).
\end{displaymath}
%
%

%
%
%
%%%%%%%%%%%%%%%%%%%%%%%%%%%%%%%%%%%%%%%%%%%%%%%%%%%%%%%%%%%%%%%%%%%%%%%%%%%%%%%%%%%%%%%%%%%%%%%%%%%%%%%%
\subsection{Neural Network Simulation}
%%%%%%%%%%%%%%%%%%%%%%%%%%%%%%%%%%%%%%%%%%%%%%%%%%%%%%%%%%%%%%%%%%%%%%%%%%%%%%%%%%%%%%%%%%%%%%%%%%%%%%%%
%
% elaborate on NN simulation
% in discussion: allows taxonomy cognitive control, automatic processing based on formal 
% discussion: how this could inform components of the cost of control
%
In order to investigate the effect of directionality for the third indirect interference during parallel processing, we implemented a synthetic neural network simulation\footnote{Simulation details are omitted due to space constraints.} identical to our earlier work \cite{CogSci_2016}. The neural network used for this simulation maps stimulus input encoded at a stimulus layer via a non-linear associative layer to  non-linear response layer. A separate task input layer encodes the current task to be performed with respect to that stimulus and projects to both the associative layer and response layer.  Units in the stimulus layer were grouped into six stimulus dimensions with three units per dimension. Similarly, units in the response layer was grouped into six response dimensions with three units per dimension. The network was trained on 12 tasks, where each task corresponds to a one-to-one mapping between a subset of three input features in a stimulus layer to a subset of three response units in an output layer. 

We then used the methods described in \citeA{CogSci_2016} to extract a bipartite task graph from single representations encoded at the associative layer. The representations associated with each task can be characterized by calculating, for each unit in the associative and output layers, the mean of its activity over all of the stimuli for a given task;  this mean pattern of activity can then be used as a representation of the task. Correlating these patterns of activity across tasks yields a task similarity matrix that can be examined separately for the associative and output layers of the network.  This can then be used to assess the extent to which different tasks rely on similar or different representation within each layer of the network. Tasks that have similar representations over the associative layer can be inferred to rely on the same input dimension — that is, they share an input component in the bipartite graph representation of the network — and tasks that are similar at the output layer can be inferred to share an output component.  Accordingly, a bipartite graph can be constructed by measuring the patterns of activity observed in the network while it performs each individual task. 

\begin{figure}[htb]
\begin{center}
  \includegraphics[width=0.45\textwidth]{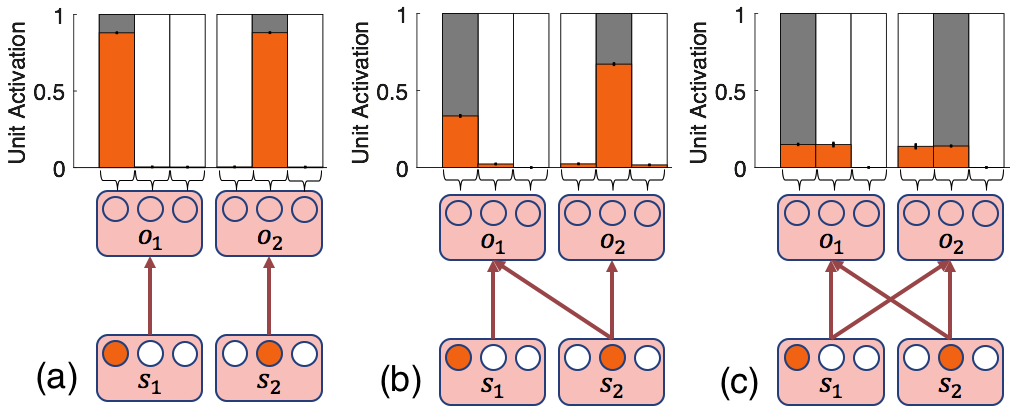}
  \caption{\small{This figure illustrates the performance of a task-pair for a given interference pattern. Each tasks maps a subset of three stimulus input units onto three response units (see text). The orange color in the bar plots indicates unit activation of response units relevant to the depicted tasks, while gray indicates desired response pattern of those units.}}
  \label{simulation}
\end{center}
\vspace{-1.5em}
\end{figure}

The extracted bipartite graph can be used to extract interference patterns between pairs of tasks (cf. Fig.~\ref{indirect}).  We use this to extract all possible learned task-pairs involving no interference case (Fig.~\ref{simulation}a), and two distinct cases of interference as shown in Fig.~\ref{simulation}b and Fig.~\ref{simulation}c from single-task representations. Fig.~\ref{simulation} shows the activation patterns of the output units for the simultaneous execution of two tasks, averaged across the patterns of all task pairs for a given interference structure. That is, no interference (Fig.~\ref{simulation}a) leads to very accurate response patterns (i.e. the current activation shown in orange) is very close to the desired activation pattern shown in grey). For the case in Fig.~\ref{simulation}b, the response pattern of task ($S_1:O_1$) is primarily impaired due to the interference arising from task ($S_2:O_1$). However, for the case of interference illustrated in Fig.~\ref{simulation}c, the response patterns for both tasks ($S_1:O_1$) and ($S_2:O_2$) are impaired as observed by the activation patterns. These simulation results reflect the influence of the directionality of interference between tasks as predicted by the proposed interaction cost.
%
%
%

%
%
%%%%%%%%%%%%%%%%%%%%%%%%%%%%%%%%%%%%%%%%%%%%%%%%%%%%%%%%%%%%%%%%%%%%%%%%%%%%%%%%%%%%%%%%%%%%%%%%%%%%%%%%
\section{General Discussion and Conclusion}
\label{sec:Conclusion}
%%%%%%%%%%%%%%%%%%%%%%%%%%%%%%%%%%%%%%%%%%%%%%%%%%%%%%%%%%%%%%%%%%%%%%%%%%%%%%%%%%%%%%%%%%%%%%%%%%%%%%%%
%%%%%%%%%%%%%%%%%%%%%%%%%%%%%%%%%%%%%%%%%%%%%%%%%%%%%%%%%%%%%%%%%%%%%%%%%%%%%%%%%%%%%%%%%%%%%%%%%%%%%%%%
%
In this study, we have introduced two new measures to determine costs associated with intensity and interaction for the demand on control. First, we quantify the intensity cost as a function of the amount of control bias that is supplementary to stimulus-specific processing in order to achieve a desired response from the network. Doing so, we formalize the probability of achieving a desired task given the stimulus, weights and biases infused to the network. Since the stimuli and weights are considered as network properties, the intensity cost to achieve desired response is defined as the amount (value) of control biases required to be injected to the input and output components of the network. The detailed analysis of intensity cost revealed an interesting optimization problem to maximize the probability of surpassing a specified activation for a given budget of resources (i.e. an upper bound on the control biases). The existence of a solution of this optimization problem implicitly reveals whether the desired objective is feasible. However, as it can be foreseen that under certain circumstances the solution does not exist due to interference between the involved processes. Such boundary conditions motivated the second metric introduced in our paper in which we formalize the interaction cost to measure the level of interactions/interference between processes by means of their type of connections and weights.

With the introduced characterization of intensity and interaction costs, it is possible to formally define whether a process is considered a reflex, automatic or controlled. Concretely, a process is considered a reflex if the underlying  weight guarantees a successful execution. In other words, a reflex can be successfully executed without any intensity or interaction costs. We assume that the execution of both controlled and automatic processes carries with it an intensity cost as some amount of control bias is needed to elicit a response. However, unlike the former, controlled processes are subject to interference and thus yield interaction costs large than zero.
  
The metrics proposed here can also be used towards further understanding cognitive effort  as well as synthetic neural networks designed to achieve goal-driven tasks. By using the intensity cost, which reveals the interrelationship between control bias and probability of achieving a desired objective, we will investigate the limitations of any given neural network architecture by allocating a budget of control bias. The intensity cost can also be used to investigate the feasibility of achieving a desired objective defined by the set of processes of interest in a network. 

In the interaction cost analysis, we have assumed that there exist a response in the output for any stimulus activation and this may not be the case for a nonlinear activation in the output components. Hence, one major research direction is the detailed analysis for the classification of processes with nonlinear activation in output. Another possible direction for future work is to further analyze the interaction cost in order to capture the properties of the overall network (not only a subset of tasks of interest). This will allow one to use the interaction cost as an objective function for network training. Another possible direction is to explore the interrelationship between intensity and interaction cost. In our work \cite{CogSci_2016}, we noticed  a fundamental trade-off between shared representations in a network and its parallel processing capability (separated representations). Intuitively, we envision this separation will decrease the interaction cost while increasing the likelihood of successful execution for a given budget of control bias. 
\section*{Acknowledgements}
We would like to thank Zahra Aminzare, Adam Charles, Jonathan Pillow and Vaibhav Srivastava for their help in the formalism of the problem.
%
%
%
%%%%%%%%%%%%%%%%%%%%%%%%%%%%%%%%%%%%%%%%%%%%%%%%%%%%%%%%%%%%%%%%%%%%%%%%%%%%%%%%%%%%%%%%%%%%%%%%%%%%%%%%
%\vspace{-1mm}
%%%%%%%%%%%%%%%%%%%%%%%%%%%%%%%%%%%%%%%%%%%%%%%%%%%%%%%%%%%%%%%%%%%%%%%%%%%%%%%%%%%%%%%%%%%%%%%%%%%%%%%%
\bibliographystyle{apacite}
\setlength{\bibleftmargin}{.125in}
\setlength{\bibindent}{-\bibleftmargin}
\bibliography{CogSciBib_2017}

\begin{thebibliography}{}

\bibitem [\protect \citeauthoryear {%
Bogacz%
, Brown%
, Moehlis%
, Holmes%
\BCBL {}\ \BBA {} Cohen%
}{%
Bogacz%
\ \protect \BOthers {.}}{%
{\protect \APACyear {2006}}%
}]{%
Bogacz_OptDecMak}
\APACinsertmetastar {%
Bogacz_OptDecMak}%
\begin{APACrefauthors}%
Bogacz, R.%
, Brown, E.%
, Moehlis, J.%
, Holmes, P.%
\BCBL {}\ \BBA {} Cohen, J\BPBI D.%
\end{APACrefauthors}%
\unskip\
\newblock
\APACrefYearMonthDay{2006}{}{}.
\newblock
{\BBOQ}\APACrefatitle {{The physics of optimal decision making: A formal
  analysis of models of performance in two-alternative forced-choice tasks.}}
  {{The physics of optimal decision making: A formal analysis of models of
  performance in two-alternative forced-choice tasks.}}{\BBCQ}
\newblock
\APACjournalVolNumPages{Psychological Review}{113}{4}{700--765}.
\PrintBackRefs{\CurrentBib}

\bibitem [\protect \citeauthoryear {%
Botvinick%
}{%
Botvinick%
}{%
{\protect \APACyear {2007}}%
}]{%
Botvinick2007}
\APACinsertmetastar {%
Botvinick2007}%
\begin{APACrefauthors}%
Botvinick, M\BPBI M.%
\end{APACrefauthors}%
\unskip\
\newblock
\APACrefYearMonthDay{2007}{}{}.
\newblock
{\BBOQ}\APACrefatitle {Conflict monitoring and decision making: Reconciling two
  perspectives on anterior cingulate function} {Conflict monitoring and
  decision making: Reconciling two perspectives on anterior cingulate
  function}.{\BBCQ}
\newblock
\APACjournalVolNumPages{Cognitive, Affective, {\&} Behavioral
  Neuroscience}{7}{4}{356--366}.
\PrintBackRefs{\CurrentBib}

\bibitem [\protect \citeauthoryear {%
Botvinick%
\ \BBA {} Braver%
}{%
Botvinick%
\ \BBA {} Braver%
}{%
{\protect \APACyear {2015}}%
}]{%
Botvinick_Braver_2015}
\APACinsertmetastar {%
Botvinick_Braver_2015}%
\begin{APACrefauthors}%
Botvinick, M\BPBI M.%
\BCBT {}\ \BBA {} Braver, T.%
\end{APACrefauthors}%
\unskip\
\newblock
\APACrefYearMonthDay{2015}{}{}.
\newblock
{\BBOQ}\APACrefatitle {{Motivation and Cognitive Control: From Behavior to
  Neural Mechanism}} {{Motivation and Cognitive Control: From Behavior to
  Neural Mechanism}}.{\BBCQ}
\newblock
\APACjournalVolNumPages{Annual Review of Psychology}{66}{1}{83--113}.
\PrintBackRefs{\CurrentBib}

\bibitem [\protect \citeauthoryear {%
Botvinick%
, Braver%
, Barch%
, Carter%
\BCBL {}\ \BBA {} Cohen%
}{%
Botvinick%
\ \protect \BOthers {.}}{%
{\protect \APACyear {2001}}%
}]{%
Botvinick_etal_2001}
\APACinsertmetastar {%
Botvinick_etal_2001}%
\begin{APACrefauthors}%
Botvinick, M\BPBI M.%
, Braver, T\BPBI S.%
, Barch, D\BPBI M.%
, Carter, C\BPBI S.%
\BCBL {}\ \BBA {} Cohen, J\BPBI D.%
\end{APACrefauthors}%
\unskip\
\newblock
\APACrefYearMonthDay{2001}{}{}.
\newblock
{\BBOQ}\APACrefatitle {Conflict monitoring and cognitive control} {Conflict
  monitoring and cognitive control}.{\BBCQ}
\newblock
\APACjournalVolNumPages{Psychological Review}{108}{3}{624--652}.
\PrintBackRefs{\CurrentBib}

\bibitem [\protect \citeauthoryear {%
Botvinick%
\ \BBA {} Cohen%
}{%
Botvinick%
\ \BBA {} Cohen%
}{%
{\protect \APACyear {2014}}%
}]{%
COGS:COGS12126}
\APACinsertmetastar {%
COGS:COGS12126}%
\begin{APACrefauthors}%
Botvinick, M\BPBI M.%
\BCBT {}\ \BBA {} Cohen, J\BPBI D.%
\end{APACrefauthors}%
\unskip\
\newblock
\APACrefYearMonthDay{2014}{}{}.
\newblock
{\BBOQ}\APACrefatitle {The Computational and Neural Basis of Cognitive Control:
  Charted Territory and New Frontiers} {The computational and neural basis of
  cognitive control: Charted territory and new frontiers}.{\BBCQ}
\newblock
\APACjournalVolNumPages{Cognitive Science}{38}{6}{1249--1285}.
\PrintBackRefs{\CurrentBib}

\bibitem [\protect \citeauthoryear {%
Botvinick%
, Cohen%
\BCBL {}\ \BBA {} Carter%
}{%
Botvinick%
\ \protect \BOthers {.}}{%
{\protect \APACyear {2004}}%
}]{%
Botvinick_etal_2004}
\APACinsertmetastar {%
Botvinick_etal_2004}%
\begin{APACrefauthors}%
Botvinick, M\BPBI M.%
, Cohen, J\BPBI D.%
\BCBL {}\ \BBA {} Carter, C\BPBI S.%
\end{APACrefauthors}%
\unskip\
\newblock
\APACrefYearMonthDay{2004}{}{}.
\newblock
{\BBOQ}\APACrefatitle {{Conflict monitoring and anterior cingulate cortex: an
  update}} {{Conflict monitoring and anterior cingulate cortex: an
  update}}.{\BBCQ}
\newblock
\APACjournalVolNumPages{Trends in Cognitive Sciences}{8}{12}{539--546}.
\PrintBackRefs{\CurrentBib}

\bibitem [\protect \citeauthoryear {%
Cohen%
, Dunbar%
\BCBL {}\ \BBA {} McClelland%
}{%
Cohen%
\ \protect \BOthers {.}}{%
{\protect \APACyear {1990}}%
}]{%
cohen1990control}
\APACinsertmetastar {%
cohen1990control}%
\begin{APACrefauthors}%
Cohen, J\BPBI D.%
, Dunbar, K.%
\BCBL {}\ \BBA {} McClelland, J\BPBI L.%
\end{APACrefauthors}%
\unskip\
\newblock
\APACrefYearMonthDay{1990}{}{}.
\newblock
{\BBOQ}\APACrefatitle {On the control of automatic processes: a parallel
  distributed processing account of the Stroop effect.} {On the control of
  automatic processes: a parallel distributed processing account of the stroop
  effect.}{\BBCQ}
\newblock
\APACjournalVolNumPages{Psychological Review}{97}{3}{332--361}.
\PrintBackRefs{\CurrentBib}

\bibitem [\protect \citeauthoryear {%
Feng%
, Schwemmer%
, Gershman%
\BCBL {}\ \BBA {} Cohen%
}{%
Feng%
\ \protect \BOthers {.}}{%
{\protect \APACyear {2014}}%
}]{%
Feng_et_al_2014}
\APACinsertmetastar {%
Feng_et_al_2014}%
\begin{APACrefauthors}%
Feng, S\BPBI F.%
, Schwemmer, M.%
, Gershman, S\BPBI J.%
\BCBL {}\ \BBA {} Cohen, J\BPBI D.%
\end{APACrefauthors}%
\unskip\
\newblock
\APACrefYearMonthDay{2014}{}{}.
\newblock
{\BBOQ}\APACrefatitle {{Multitasking vs. Multiplexing: Toward a normative
  account of limitations in the simultaneous execution of control-demanding
  behaviors}} {{Multitasking vs. Multiplexing: Toward a normative account of
  limitations in the simultaneous execution of control-demanding
  behaviors}}.{\BBCQ}
\newblock
\APACjournalVolNumPages{Cognitive, Affective, \& Behavioral
  Neuroscience}{14}{1}{129-146}.
\PrintBackRefs{\CurrentBib}

\bibitem [\protect \citeauthoryear {%
Koechlin%
\ \BBA {} Summerfield%
}{%
Koechlin%
\ \BBA {} Summerfield%
}{%
{\protect \APACyear {2007}}%
}]{%
koechlin2007information}
\APACinsertmetastar {%
koechlin2007information}%
\begin{APACrefauthors}%
Koechlin, E.%
\BCBT {}\ \BBA {} Summerfield, C.%
\end{APACrefauthors}%
\unskip\
\newblock
\APACrefYearMonthDay{2007}{}{}.
\newblock
{\BBOQ}\APACrefatitle {An information theoretical approach to prefrontal
  executive function} {An information theoretical approach to prefrontal
  executive function}.{\BBCQ}
\newblock
\APACjournalVolNumPages{Trends in Cognitive Sciences}{11}{6}{229--235}.
\PrintBackRefs{\CurrentBib}

\bibitem [\protect \citeauthoryear {%
Kurzban%
, Duckworth%
, Kable%
\BCBL {}\ \BBA {} Myers%
}{%
Kurzban%
\ \protect \BOthers {.}}{%
{\protect \APACyear {2013}}%
}]{%
kurzban2013opportunity}
\APACinsertmetastar {%
kurzban2013opportunity}%
\begin{APACrefauthors}%
Kurzban, R.%
, Duckworth, A.%
, Kable, J\BPBI W.%
\BCBL {}\ \BBA {} Myers, J.%
\end{APACrefauthors}%
\unskip\
\newblock
\APACrefYearMonthDay{2013}{}{}.
\newblock
{\BBOQ}\APACrefatitle {An opportunity cost model of subjective effort and task
  performance.} {An opportunity cost model of subjective effort and task
  performance.}{\BBCQ}
\newblock
\APACjournalVolNumPages{The Behavioral and Brain Sciences}{36}{6}{661--679}.
\PrintBackRefs{\CurrentBib}

\bibitem [\protect \citeauthoryear {%
Muraven%
, Tice%
\BCBL {}\ \BBA {} Baumeister%
}{%
Muraven%
\ \protect \BOthers {.}}{%
{\protect \APACyear {1998}}%
}]{%
muraven1998self}
\APACinsertmetastar {%
muraven1998self}%
\begin{APACrefauthors}%
Muraven, M.%
, Tice, D\BPBI M.%
\BCBL {}\ \BBA {} Baumeister, R\BPBI F.%
\end{APACrefauthors}%
\unskip\
\newblock
\APACrefYearMonthDay{1998}{}{}.
\newblock
{\BBOQ}\APACrefatitle {Self-control as a limited resource: Regulatory depletion
  patterns} {Self-control as a limited resource: Regulatory depletion
  patterns}.{\BBCQ}
\newblock
\APACjournalVolNumPages{Journal of Personality and Social
  Psychology}{74}{3}{774-789}.
\PrintBackRefs{\CurrentBib}

\bibitem [\protect \citeauthoryear {%
Musslick%
\ \protect \BOthers {.}}{%
Musslick%
\ \protect \BOthers {.}}{%
{\protect \APACyear {2016}}%
}]{%
CogSci_2016}
\APACinsertmetastar {%
CogSci_2016}%
\begin{APACrefauthors}%
Musslick, S.%
, Dey, B.%
, {\"O}zcimder, K.%
, Patwary, M\BPBI M\BPBI A.%
, Willke, T\BPBI L.%
\BCBL {}\ \BBA {} Cohen, J\BPBI D.%
\end{APACrefauthors}%
\unskip\
\newblock
\APACrefYearMonthDay{2016}{}{}.
\newblock
{\BBOQ}\APACrefatitle {{Controlled vs. Automatic Processing: A Graph-Theoretic
  Approach to the Analysis of Serial vs. Parallel Processing in Neural Network
  Architectures}} {{Controlled vs. Automatic Processing: A Graph-Theoretic
  Approach to the Analysis of Serial vs. Parallel Processing in Neural Network
  Architectures}}.{\BBCQ}
\newblock
\BIn{} \APACrefbtitle {{Proceedings of the 38th Annual Conference of the
  Cognitive Science Society}} {{Proceedings of the 38th Annual Conference of
  the Cognitive Science Society}}\ (\BPGS\ 1547--1552).
\newblock
\APACaddressPublisher{Philadelphia, PA}{}.
\PrintBackRefs{\CurrentBib}

\bibitem [\protect \citeauthoryear {%
Ortega%
\ \BBA {} Braun%
}{%
Ortega%
\ \BBA {} Braun%
}{%
{\protect \APACyear {2013}}%
}]{%
ortega2013thermodynamics}
\APACinsertmetastar {%
ortega2013thermodynamics}%
\begin{APACrefauthors}%
Ortega, P\BPBI A.%
\BCBT {}\ \BBA {} Braun, D\BPBI A.%
\end{APACrefauthors}%
\unskip\
\newblock
\APACrefYearMonthDay{2013}{}{}.
\newblock
{\BBOQ}\APACrefatitle {Thermodynamics as a theory of decision-making with
  information-processing costs} {Thermodynamics as a theory of decision-making
  with information-processing costs}.{\BBCQ}
\newblock
\APACjournalVolNumPages{Proceedings of Royal Society A}{469}{2153}{20120683}.
\PrintBackRefs{\CurrentBib}

\bibitem [\protect \citeauthoryear {%
Posner%
\ \BBA {} Snyder%
}{%
Posner%
\ \BBA {} Snyder%
}{%
{\protect \APACyear {1975}}%
}]{%
posnerr}
\APACinsertmetastar {%
posnerr}%
\begin{APACrefauthors}%
Posner, M\BPBI I.%
\BCBT {}\ \BBA {} Snyder, C\BPBI R\BPBI R.%
\end{APACrefauthors}%
\unskip\
\newblock
\APACrefYearMonthDay{1975}{}{}.
\newblock
{\BBOQ}\APACrefatitle {Attention and Cognitive Control} {Attention and
  cognitive control}.{\BBCQ}
\newblock
\BIn{} R\BPBI L.~Solso\ (\BED), \APACrefbtitle {{Information Processing and
  Cognition: The Loyola Symposium}} {{Information Processing and Cognition: The
  Loyola Symposium}}\ (\BPG~55-85).
\newblock
\APACaddressPublisher{}{Lawrence Erlbaum}.
\PrintBackRefs{\CurrentBib}

\bibitem [\protect \citeauthoryear {%
Shenhav%
, Botvinick%
\BCBL {}\ \BBA {} Cohen%
}{%
Shenhav%
\ \protect \BOthers {.}}{%
{\protect \APACyear {2013}}%
}]{%
shenhav2013expected}
\APACinsertmetastar {%
shenhav2013expected}%
\begin{APACrefauthors}%
Shenhav, A.%
, Botvinick, M\BPBI M.%
\BCBL {}\ \BBA {} Cohen, J\BPBI D.%
\end{APACrefauthors}%
\unskip\
\newblock
\APACrefYearMonthDay{2013}{}{}.
\newblock
{\BBOQ}\APACrefatitle {The expected value of control: an integrative theory of
  anterior cingulate cortex function} {The expected value of control: an
  integrative theory of anterior cingulate cortex function}.{\BBCQ}
\newblock
\APACjournalVolNumPages{Neuron}{79}{2}{217--240}.
\PrintBackRefs{\CurrentBib}

\bibitem [\protect \citeauthoryear {%
Shenhav%
\ \protect \BOthers {.}}{%
Shenhav%
\ \protect \BOthers {.}}{%
{\protect \APACyear {2017}}%
}]{%
Shenav_PrePrint_ARP}
\APACinsertmetastar {%
Shenav_PrePrint_ARP}%
\begin{APACrefauthors}%
Shenhav, A.%
, Musslick, S.%
, Lieder, F.%
, Kool, W.%
, Griffiths, T\BPBI L.%
, Cohen, J\BPBI D.%
\BCBL {}\ \BBA {} Botvinick, M\BPBI M.%
\end{APACrefauthors}%
\unskip\
\newblock
\APACrefYearMonthDay{2017}{}{}.
\newblock
\APACrefbtitle {Toward a Rational and Mechanistic Account of Mental Effort.}
  {Toward a rational and mechanistic account of mental effort.}
\newblock
\APACrefnote{submitted to Annual Reviews of Neuroscience}
\PrintBackRefs{\CurrentBib}

\bibitem [\protect \citeauthoryear {%
Shiffrin%
\ \BBA {} Schneider%
}{%
Shiffrin%
\ \BBA {} Schneider%
}{%
{\protect \APACyear {1977}}%
}]{%
shiffrin1977controlled}
\APACinsertmetastar {%
shiffrin1977controlled}%
\begin{APACrefauthors}%
Shiffrin, R\BPBI M.%
\BCBT {}\ \BBA {} Schneider, W.%
\end{APACrefauthors}%
\unskip\
\newblock
\APACrefYearMonthDay{1977}{}{}.
\newblock
{\BBOQ}\APACrefatitle {{Controlled and automatic human information processing:
  II. Perceptual learning, automatic attending and a general theory}}
  {{Controlled and automatic human information processing: II. Perceptual
  learning, automatic attending and a general theory}}.{\BBCQ}
\newblock
\APACjournalVolNumPages{Psychological Review}{84}{2}{127-190}.
\PrintBackRefs{\CurrentBib}

\end{thebibliography}
%%%%%%%%%%%%%%%%%%%%%%%%%%%%%%%%%%%%%%%%%%%%%%%%%%%%%%%%%%%%%%%%%%%%%%%%%%%%%%%%%%%%%%%%%%%%%%%%%%%%%%%%
%
%
%
\end{document}